\documentclass{elsart}
\usepackage{elsfix}
\usepackage{epsfig}

\begin{document}
\begin{frontmatter}
\title{Local Magnetic Susceptibility of the Positive Muon in the Quasi
1D S=1/2 Antiferromagnet KCuF$_3$
\\ \fontfamily{cmtt}\selectfont{(to be published in Physica B)}}
\author[UBC]{\underline{J. Chakhalian}},
 \author[UBC,CIAR,TRIUMF]{R.F. Kiefl},
\author[UBC]{R. Miller},
\author[LANL]{S.R. Dunsiger},
 \author[LANL]{G. Morris},
 \author[TRIUMF]{S. Kreitzman},
\author[TRIUMF,CHEM]{W.A. MacFarlane},
 \author[SFU]{J. Sonier},
 \author[CHALM]{S. Eggert},
 \author[UBC,CIAR,BOSTON]{I. Affleck}
\and \author[CHIBA]{I. Yamada}

\address[UBC]{Department of Physics and Astronomy, The University of British
Columbia, Vancouver BC, V6T 1Z1, Canada}
\address[CIAR]{Canadian Institute for Advanced Research}
\address[TRIUMF]{TRIUMF, 4004 Wesbrook Mall, Vancouver,  BC, V6T 2A3,
Canada}
\address[LANL]{Los Alamos National Lab,
 MST-10, MS K764, Los Alamos, NM 87545, USA}
\address[SFU] {Department of Physics, Simon Fraser University, Burnaby,
British Columbia, V5A 1S6, Canada}
\address[CHEM]{Chemistry Department, University of British Columbia, Vancouver, 
Canada}
\address[CHALM]{Institute of Theoretical Physics, Chalmers University of Technology and
G\"{o}teborg University, S412 96 G\"{o}teborg, Sweden}
\address[BOSTON]{Physics Department, Boston University,
 590 Commonwealth Ave., Boston, MA02215, USA}
\address[CHIBA] {Department of Physics, Faculty of Science, Chiba University,
Yayoi-cho, Inage-ku, Chiba 263-8522, Japan}

\begin{abstract} 

We report muon spin rotation measurements of
the local magnetic susceptibility around a positive muon in the paramagnetic
state of   the  quasi one-dimensional spin 1/2 antiferromagnet KCuF$_3$.
Signals from two distinct sites  are resolved which have a temperature dependent
frequency shift which is different than the 
magnetic susceptibility.  This difference   is attributed to a muon 
induced perturbation of the spin 1/2 chain.  

\end{abstract}
\end{frontmatter}

\underline{Corresponding author:} \\
Dr. Jacques Chakhalian, Department of Physics and Astronomy,
The University of British Columbia, Vancouver, British Columbia,
V6T 1Z1, Canada. 
Tel. 604-222-1047, FAX: 604-222-4710, e-mail: jacques@triumf.ca

Novel magnetic effects are predicted for a non-magnetic impurity in a
one dimensional spin 1/2 antiferromagnetic chain
\cite{ref:amf1,ref:amf2,eggert}. In particular, at low temperatures
the magnetic susceptibility in the region of a perturbed link is
expected to differ dramatically from the uniform bulk susceptibility
Furthermore, the effects of such a perturbation propagate far along
the chain and differ depending on whether the perturbation is link or
site symmetric.  The effect is closely related to Kondo screening of a
magnetic impurity in a metal, and arises in part because of the
gapless spectrum of excitations which characterizes a Heisenberg spin
1/2 chain.  Although truly one dimensional spin 1/2 chains have no
long range ordering above $T=0$, real materials always exhibit 3D
N$\acute{e}$el ordering due the finite interchain coupling, $J_\perp$.
Nevertheless the one dimensional properties can be studied down to low
temperatures ($T\ll J$) in quasi one dimensional systems where
$J_\perp \ll J_\parallel$.

A $\mu$SR experiment is an ideal way to test such  ideas since 
the muon acts as both the impurity and the 
probe of the local magnetic susceptibility. We anticipate that the
positively charged muon will distort the crystal lattice, thereby
altering  the exchange coupling between the magnetic ions in
the vicinity of the muon. The resulting modification of the local
susceptibility will  be reflected in the muon frequency shift.

In this paper we report muon spin rotation measurements on a single
crystal of KCuF$_3$, which is a well known quasi 1-D Heisenberg S=1/2
antiferromagnet \cite{xtall1}, \cite{xtall2}.  We find evidence for
two magnetically inequivalent F$\mu$F centers in which the muon is
hydrogen bonded to two neighboring F$^-$ ions implying a large lattice
distortion.  The raw frequency shifts are opposite for the two sites
and they display temperature dependence which is distinctly different
than the bulk magnetic susceptibility ($\chi$).  These effects are
attributed to a muon induced perturbation of the local spin
susceptibility.
 
\begin{figure} 
   \epsfig{figure=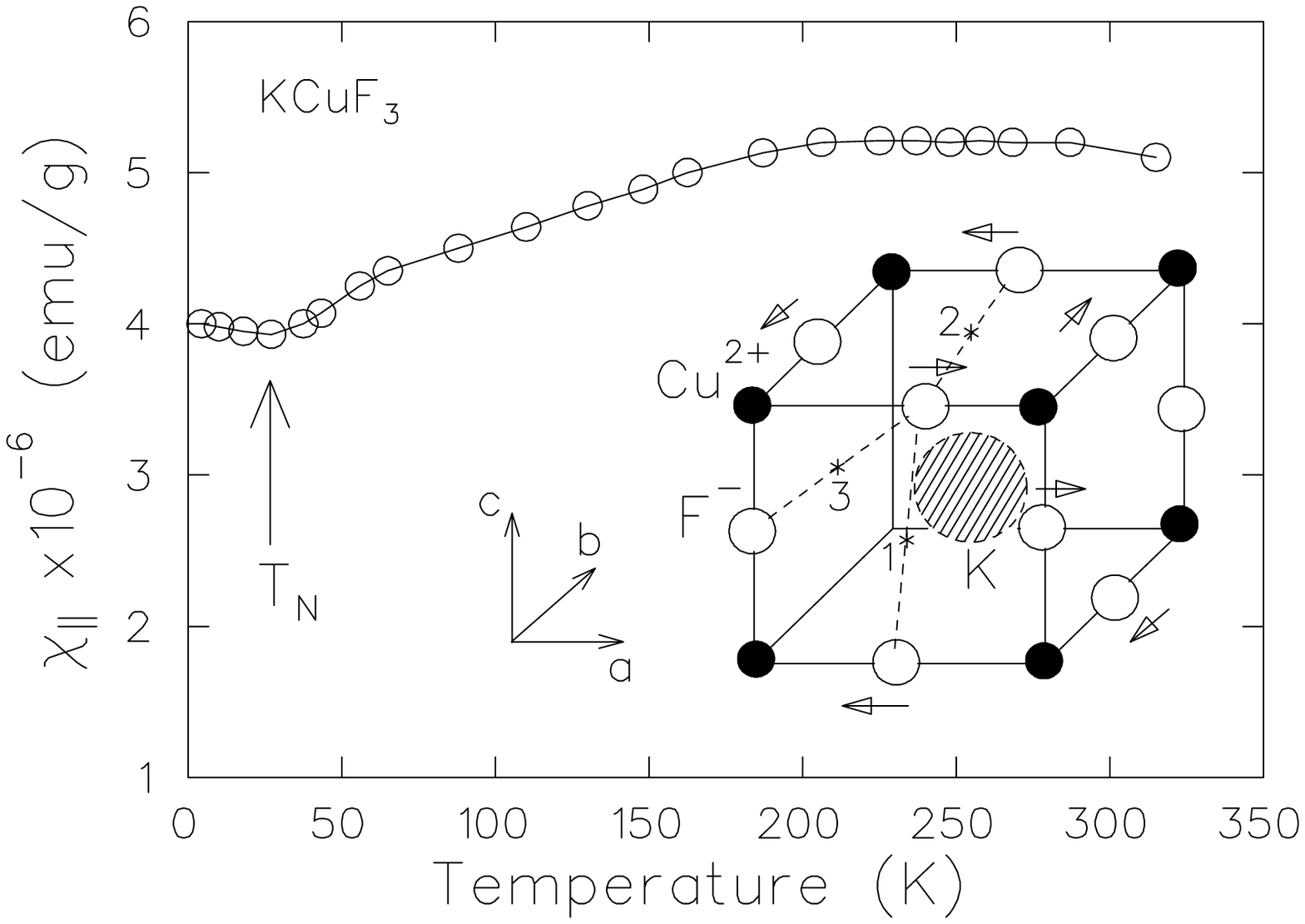, width=15cm}
\caption{ Magnetic susceptibility of KCuF$_3$ measured in a SQUID magnetometer in 
 the applied magnetic field of 1 T.  An inset shows a pseudo unit cell  of  KCuF$_3$. 
The displaced F$^-$ ions are indicated by arrows. 
Possible muon sites are indicated 
by the ($\ast$) symbol.}
    \label{figure1}
 \end{figure}

KCuF$_3$ has a tetragonal crystal structure with lattice parameters
$c=3.914$ $\rm \AA$ and $a=4.126$ $\rm \AA$ at 10 K (see inset in Fig.
\ref{figure1}).  The structure is similar to a perovskite. However, a
Jahn-Teller distortion in the $a-b$ plane causes F$^-$ ions in the
$a-b$ plane to be displaced slightly away from the edge center by 0.31
$\rm \AA$ \cite{xtall3}.  The magnetic properties of KCuF$_3$ arise
from the S=1/2 Cu$^{2+}$ ions which are almost perfectly Heisenberg
coupled but with very different coupling strengths for spins along the
$c$-axis versus in the $a-b$ plane. The ratio between the interchain
and intrachain coupling constants $J_\perp/J_\parallel=0.01$ with
$J_\parallel=190$ K so the system is very one dimensional.  There are
two polytypes (a and d) with slightly different arrangements of F$^-$
ions and N$\acute{e}$el temperatures of 39.3 K and 22.7 K respectively
\cite{xtall2}.  The crystal used in this experiment was polytype a as
shown in Fig.1.  Recent ZF-$\mu$SR results indicate the magnetic
transition in polytype a is first order\cite{mazzoli}.

 All the measurements were performed at the M20 beamline at TRIUMF
 which delivers nearly 100$\%$ spin polarized positive muons with a
 mean momentum of 28 MeV/c.  The muon spin polarization was rotated
 perpendicular to the axis of the superconducting solenoid and muon
 beam direction.  The magnitude of the applied magnetic field $H=1.45$
 T was chosen to provide a balance between the magnitude of the
 frequency shift which increases with field and the amplitude of the
 $\mu$SR signal which eventually diminishes with increasing field due
 to the finite timing resolution of the detectors.  The transverse
 field precession measurements were all performed with a special
 cryostat insert which allows spectra to be taken on the sample and on
 a reference material simultaneously \cite{nikko}.

\begin{figure}
   \epsfig{figure=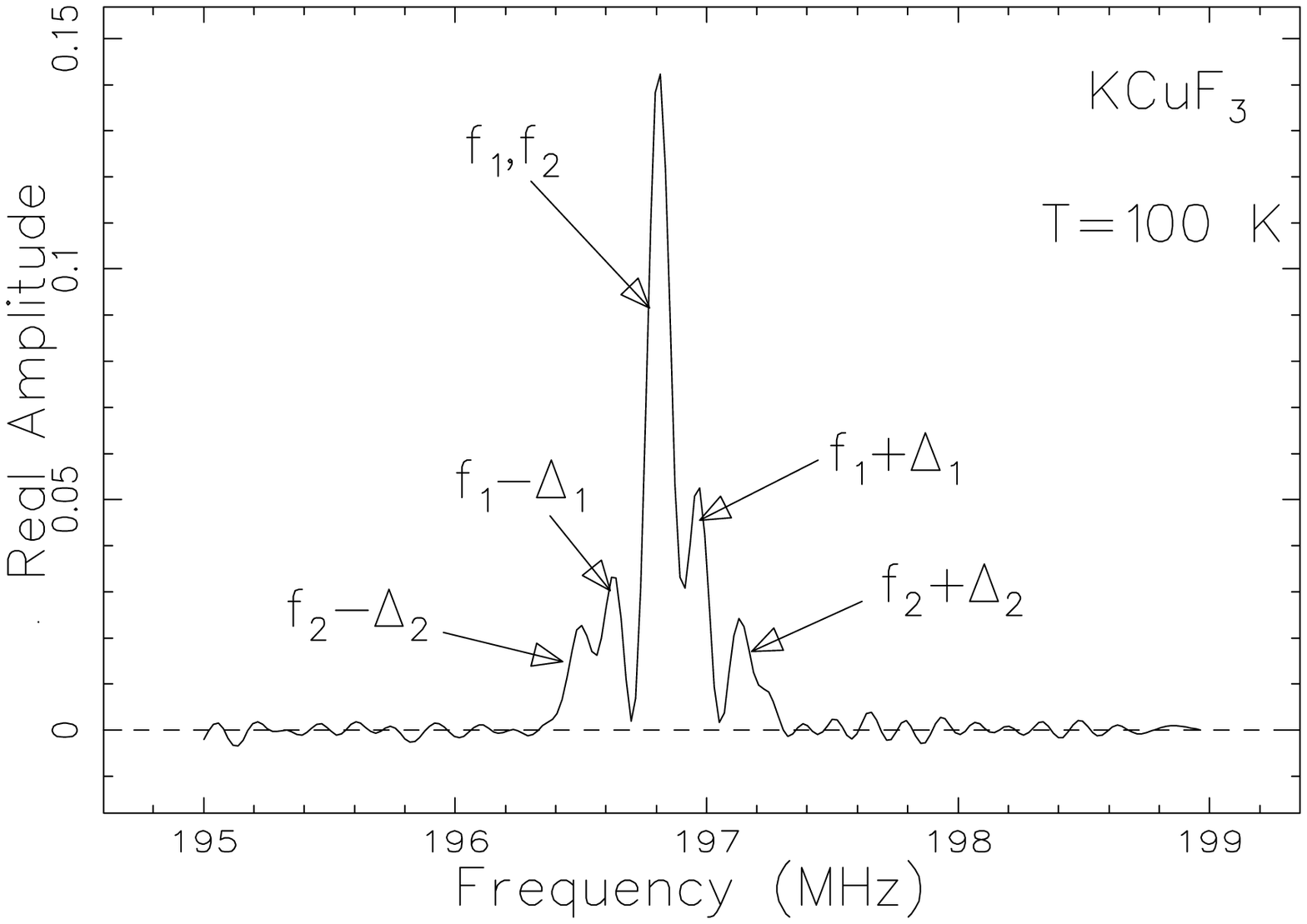, width=15cm}
\caption{Fourier transform of the $\mu$SR time spectrum
measured in a field of 1.45 T applied along the c-axis.}
   \label{figure2}
\end{figure}
 
Figure \ref{figure2} shows frequency spectra at 100 K which was
obtained by fast fourier transforming the muon spin precession signal,
which is analogous to the free induction decay in an NMR experiment.
All the frequency shift measurements were taken with the external
magnetic field applied along the c-axis. Near room temperature one
observes a single narrow line, which is attributed to fast muon
diffusion whereby the dipolar interactions with nuclear magnetic
moments are motionally averaged. Below room temperature the line
broadens and develops clear splittings as shown in Fig. 2.  Such
splittings are attributed to the large $^{19}$F nuclear moments and
provide important information on the muons site and the symmetry of
the perturbation that the muon induces.  The observed splittings are
characteristic of a static F$\mu$F center in which the positive muon
forms a collinear ionic bond between two F$^-$ ions\cite{jess}.  The
presence of the F$\mu$F center was also confirmed with measurements in
zero applied field. Figure 3 shows the characteristic muon oscillation
in zero applied field for F$\mu$F in KCuF$_3$.  The curve is a fit to
the polarization signal generated from the spin Hamiltonian for
F$\mu$F with a muon-$^{19}$F nuclear dipolar coupling
$\nu_d=\gamma_\mu\gamma_F/r^3=0.216$ MHz where $r$ is muon-$^{19}$F
distance.  This value of $\nu_d$ is typical of that seen in many
compounds containing fluorine and implies a F-F separation of 2.38
\AA\ which is about twice the ionic radius of the F$^-$
ion\cite{jess}. Similar ZF spectra have recently been reported in
polycrystalline KCuF$_3$ \cite{mazzoli}.  In a high transverse
magnetic field one expects that each F$\mu$F center will give rise to
a triplet of lines with an amplitude ratio of $1:2:1$ and
corresponding frequencies:
\begin{eqnarray}
\nu^- &=& \gamma_\mu B - \nu_d(1-3\cos^2\theta) \nonumber \\
\nu^0 &=& \gamma_\mu B \nonumber \\
\nu^+ &=& \gamma_\mu B + \nu_d(1-3\cos^2\theta)  
\end{eqnarray}
where $B$ is the local magnetic field at the muon with no  contribution from the 
$^{19}$F nuclear moments, and   $\theta$ is the angle between the 
 magnetic field and the F$\mu$F bond axis.
Note from  the spectrum at 100 K in 
Fig. \ref{figure2} that four satellite lines are well resolved, implying 
two magnetically inequivalent
F$\mu$F centers  with two distinct values of $\theta$.   
The central lines are  unaffected by the nuclear dipolar coupling and therefore
are not resolved in the spectrum. 
 
\begin{figure}
   \epsfig{figure=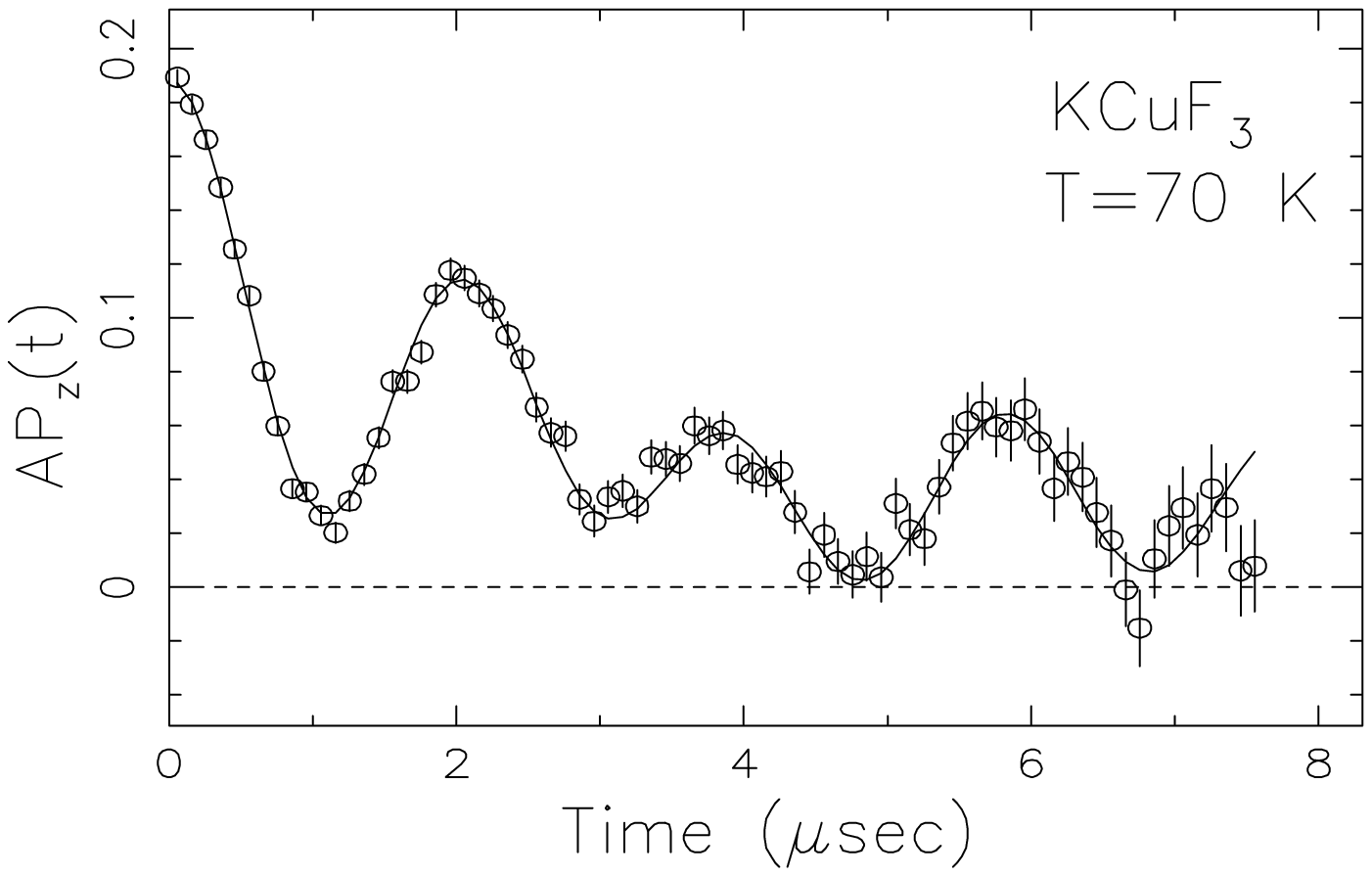, width=15cm}
\caption{Time evolution of the muon polarization 
in a zero applied magnetic field. The signal is characteristic
of the muon-$^{19}$F nuclear dipolar interaction  in an F$\mu$F center.}
   \label{figure3}
 \end{figure}

 Good fits to all the data between 50 K and 200 K were obtained with
 the above model assuming two static F$\mu$F centers with satellite
 splittings of $\Delta_1=0.17(1)$ MHz and $\Delta_2=0.30(1)$ MHz for
 sites 1 and 2 respectively.  These splittings are slightly less than
 one would expect from the face center positions (sites 1 and 2 in
 Fig. 1) assuming that the angle between field (or c-axis) and the F-F direction is
 unchanged by the muon.  In this case we would expect dipolar
 splittings of approximately $\nu_d$ and $1.9\nu_d$; whereas, the
 observed splittings are 0.8$\nu_d$ and $1.4\nu_d$ respectively.
 Diagonal sites (site 3) are possible but the splittings for these
 sites should be about $0.5\nu_d$ and $\nu_d$.  We are led to the
 conclusion that F-F internuclear direction rotates slightly by the
 presence of the muon.  In retrospect this is reasonable considering
 that the Jahn-Teller distortion displaces the F$^-$ ions in the $a-b$
 plane so that the Cu-F bonds are not of equal strength.  Therefore we
 attribute the two signals to muons at sites 1 and 2 in Fig. 1. The
 large contraction of the F-F distance is typical of F$\mu$F centers
 seen in other F containing compounds ionic fluorides\cite{jess}.
 However, KCuF$_3$ is unusual in that the muon also produces a small
 rotation of the F-F internuclear direction.  Such a large lattice
 distortion should produce a significant perturbation of the exchange
 coupling between the nearest neighbor Cu$^{2+}$ spins.

\begin{figure}
   \epsfig{figure=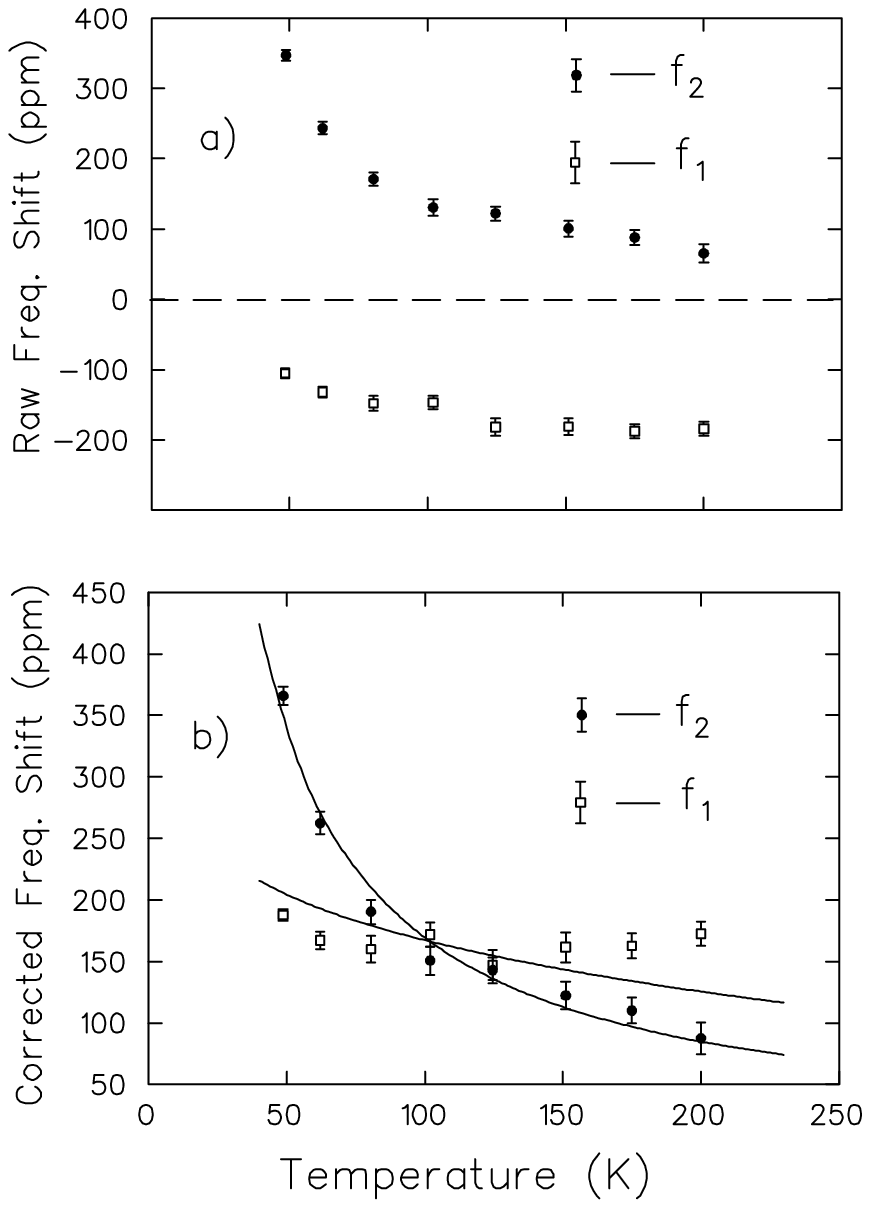, width=15cm}
\caption{(a) Temperature dependence of the raw 
muon frequency  shifts at two interstitial sites in a  magnetic field 
H=$1.45$ T. (b) Frequency shifts for sites 1 and 2 subtracting 
the dipolar fields from all Cu$^{2+}$  moments
other than the four nearest neighbors}
   \label{figure4}
 \end{figure}

 Measurements of the precession frequency signal in the sample and a
 reference material (Ag) were taken simultaneously. This eliminates
 many systematic effects, such as field drift, which are important
 when the frequency shifts are small. After correcting for the
 temperature independent Knight shift in Ag (+94 ppm) \cite{schenck}
 and the small difference in field between the reference and sample
 (22 ppm) we obtain the frequency shifts for the two sites shown in
 Figure \ref{figure4}.

 Note that the raw frequency shifts of $f_1$ and $f_2$ are similar but
 have opposite sign. This difference in sign is attributed to site
 dependent dipolar field from the polarized Cu$^{2+}$ moments. For
 example if one subtracts a calculated dipolar field from all
 Cu$^{2+}$ moments except the four nearest neighbor Cu$^{2+}$ assuming
 these more distant moments are polarized according to the bulk $\chi$
 shown in Fig. 1, then one obtains the corrected frequency shifts
 shown in Fig.  4b. As may be inferred from Figs. 4a and 4b this
 correction is large and negative for site 1 and almost zero at site
 2.  The corrected frequency shifts in Fig. 4b are then both positive
 and originate from the local dipolar field from the four nearest
 neighbor Cu moments plus the contact interaction.  Note that the
 temperature dependence of the raw and corrected shifts are somewhat
 different, which is due to the fact that the size of the corrections
 are different and scale with the bulk susceptibility in Fig. 1.
 Clearly the dipolar corrections to the frequency shift depend on the
 site. However, given the large deviations between the raw frequency
 shifts and the bulk $\chi$ (Fig. 1) the local $\chi$ must also be
 very different.  In particular, the bulk $\chi$ in a spin 1/2 chain
 peaks at around $J$ and decreases at lower temperatures due to short
 range AF correlations.  This is clearly not the case for $f_2$ where
 the magnitude of the shift increases dramatically below 200 K.  The
 temperature dependence of $f_1$ on the other hand is somewhat weaker,
 but still quite different from the pure susceptibility.
 This behavior is in agreement with the theoretical
 predictions\cite{ref:amf1,ref:amf2,eggert,seb1,seb2}.  The solid
 lines in Fig. 4b show a quantitative comparison with the theoretical
 calculations\cite{ref:amf1,ref:amf2,eggert,seb1,seb2} assuming a
 completely broken link (location 1) and two completely broken links
 (location 2), respectively.  Here the muon has been assumed to feel
 the local magnetic moment of the nearest copper atoms via a contact
 interaction of unknown strength.  There are no other adjustable
 parameters in this fit.  The overall agreement is rather convincing,
 but some deviations should be expected since we assumed earlier that
 all Cu-atoms away from the muon have the same dipole moment, which is
 a simplification since an impurity in a one dimensional system will
 affect many magnetic sites in the
 chain\cite{ref:amf1,ref:amf2,eggert,seb1,seb2}.
 In summary, the local magnetic susceptibility around the muon in
 quasi 1D S=1/2 antiferromagnetic chain compound KCuF$_3$ has been
 investigated using $\mu$SR. Signals from two distinct sites are
 identified and shown to have the local magnetic susceptibilities
 which are different from each other and also different than the bulk
 $\chi$. The theoretical fits capture the effect of muon perturbation
 rather well.  These results confirm the high sensitivity of one
 dimensional spin 1/2 chain compounds to impurities.

This research was supported by  NSERC, CIAR and the Centre for Materials and
Molecular Research at TRIUMF. 
 

\end{document}